\documentclass[prb,twocolumn,superscriptaddress,showpacs,aps]{revtex4-1}

\usepackage{color}
\usepackage{graphicx}
\usepackage{bm}
\usepackage{amsmath}

\begin{document}

\newcommand{\Tc}{T_{\text c}}
\newcommand{\Tn}{T_N}
\newcommand{\Hcii}{H_{\text c2}}
\newcommand{\Tm}{TmNi$_2$B$_2$C}
\newcommand{\Lu}{LuNi$_2$B$_2$C}
\newcommand{\Ce}{CeCoIn$_5$}

\title{Field Dependence of the Superconducting Basal Plane Anisotropy of \Tm}

\author{P.~Das}
\affiliation{Department of Physics, University of Notre Dame, Notre Dame, Indiana 46556, USA}

\author{J.~M.~Densmore}
\altaffiliation{Current address: Energetic Materials Center, Lawrence Livermore National Laboratory, Livermore, California 94550}
\affiliation{U. S. Army Research Laboratory, Aberdeen Proving Ground, Maryland 21005-5066}

\author{C.~Rastovski}
\affiliation{Department of Physics, University of Notre Dame, Notre Dame, Indiana 46556, USA}

\author{K.~J.~Schlesinger}
\altaffiliation{Current address: Department of Physics, University of California, Santa Barbara, California, 93106}
\affiliation{Department of Physics, University of Notre Dame, Notre Dame, Indiana 46556, USA}

\author{M.~Laver}
\affiliation{Laboratory for Neutron Scattering, Paul Scherrer Insitut, CH-5232 Villigen, Switzerland}
\affiliation{Materials Research Division, Ris\o \ DTU, Technical University of Denmark, DK-4000 Roskilde, Denmark}
\affiliation{Nano Science Center, Niels Bohr Institute, University of Copenhagen, DK-2100 Copenhagen, Denmark}

\author{C.~D.~Dewhurst}
\affiliation{Institut Laue-Langevin, 6 Rue Jules Horowitz, F-38042 Grenoble, France}

\author{K.~Littrell}
\affiliation{Oak Ridge National Laboratory, Oak Ridge, Tennessee 37831-6393, USA}

\author{S.~L.~Bud'ko}
\author{P.~C.~Canfield}
\affiliation{Ames Laboratory and Department of Physics and Astronomy, Iowa State University, Ames, Iowa 50011, USA}

\author{M.~R.~Eskildsen}
\email{eskildsen@nd.edu}
\affiliation{Department of Physics, University of Notre Dame, Notre Dame, Indiana 46556, USA}

\date{\today}

\begin{abstract}
The superconductor \Tm \ possesses a significant four-fold basal plane anisotropy, leading to a square Vortex Lattice (VL) at intermediate fields. However, unlike other members of the borocarbide superconductors, the anisotropy in \Tm \ appears to decrease with increasing field, evident by a reentrance of the square VL phase. We have used Small Angle Neutron Scattering measurements of the VL to study the field dependence of the anisotropy. Our results provide a direct, quantitative measurement of the decreasing anisotropy. We attribute this reduction of the basal plane anisotropy to the strong Pauli paramagnetic effects observed in \Tm \ and the resulting expansion of vortex cores near $\Hcii$.
\end{abstract}

\pacs{74.25.Uv,74.70.Dd,61.05.fg}

\maketitle

\section{Introduction}
The vortex lattice (VL) symmetry and orientation in clean type-II superconductors depends sensitively on the host material anisotropy, vortex density and temperature, frequently leading to rich phase diagrams. As a result, VL studies can be used as a sensitive probe of the anisotropy of the superconducting state.

In superconductors with a sufficient four-fold basal plane anisotropy, either due to the pairing symmetry or the Fermi velocity, the VL undergoes a successive series of generic symmetry and orientational transitions as the vortex density is increased:\cite{RefWorks:751,RefWorks:754} At low fields a distorted hexagonal VL is observed, oriented with the unit cell diagonal along a crystalline high symmetry direction. As the field is increased the VL undergoes a first-order reorientation and symmetry transition to a rhombic phase with the unit cell diagonal rotated $45^\circ$ with respect to the hexagonal VL phase. Finally, upon further increase of the field, the rhombic VL continuously transforms into a square symmetry. The transitions are driven by the growing importance of the four-fold anisotropy of the vortex-vortex interaction as the vortex density increases, explaining why further changes of the VL structure are usually not observed once the square phase has been reached. There exists, however, two striking exceptions to this behavior as seen in the superconductors \Tm \ and \Ce.\cite{RefWorks:38,RefWorks:22,RefWorks:5,RefWorks:331,RefWorks:1} In both of these materials, the VL undergoes the normal progression of symmetry transitions described above at low fields. However, the square phase VL is found to be reentrant, and the VL undergoes the same sequence of transitions but in the reverse order as the field is further increased. This indicates a reduction of the superconducting basal plane anisotropy in these materials at high fields and is the main objective of this report.

In both \Tm \ and \Ce \ the superconducting state is strongly affected by Pauli paramagnetic effects.\cite{RefWorks:8,RefWorks:6,RefWorks:5,RefWorks:1} Briefly, there is a significant spin-polarization of the unpaired quasi-particles in the vortex cores, resulting in an increased amplitude of the modulation of the magnetic field.\cite{RefWorks:756,RefWorks:805,RefWorks:804,Aoyama} With increasing field the vortex cores are also predicted to expand and become more isotropic, leading to the reverse sequence of VL transitions.\cite{IchiokaPC,RefWorks:867}

Here we present the results of small-angle neutron scattering (SANS) experiments to directly measure the evolution of the basal plane anisotropy in the high-field square, rhombic and hexagonal VL phases. This is possible by measuring a large number of higher order VL reflections in a manner analogous to our previous study of non-magnetic \Lu \ (no Pauli paramagnetic effects and no reentrance of the square VL phase).\cite{RefWorks:3} Our measurements allow a quantitative determination of the four-fold basal plane anisotropy and show a monotonic decrease with increasing field.

\section{Experimental}
The sample was a single crystal of {\Tm} of mass 387~mg and dimensions $9.0 \times 8.0 \times 1.0$~mm$^3$ with the $c$ axis along the thin direction, grown by a high temperature flux method and using isotopically enriched $^{11}$B to reduce neutron absorption.\cite{RefWorks:397} \Tm \ has a superconducting critical temperature $\Tc = 11$~K and a N\'{e}el temperature $\Tn = 1.5$~K below which the Tm moments order antiferromagnetically (AFM).\cite{RefWorks:269,RefWorks:870,RefWorks:463,RefWorks:868} The sample was mounted on an aluminium plate with the crystalline $a$ axis vertical. The $c$ axis is horizontal and rotated by an angle $\Omega$ with respect to the applied magnetic field and the incoming neutrons as shown in the inset to Fig.~\ref{Fig1}(a). The rotation ($\Omega$) can favor a single VL domain orientation, thus reducing complications to the data analysis resulting from overlapping peaks from different domains while measuring higher orders of Bragg reflections. The SANS experiment was carried out at the 30~m NG7 instrument at the NIST Center for Neutron Research, using a neutron wavelength $\lambda_{\text n} = 0.55$~nm and a spread $\Delta\lambda_{\text n}/\lambda_{\text n} = 11{\%}$~FWHM. A horizontal field cryomagnet was used to reach the desired fields and temperatures. Measurements were done at a temperature of $1.6$~K and in a field range $0.2 \leq \mu_0 H \leq 0.6$~T. Preliminary measurements were also carried out at the NG-2 SANS instrument at Oak Ridge National Laboratory and at the D11 SANS at Institut Laue-Langevin.

Comparing the field dependence of the VL form factor (see Sect.~III) for the first-order reflections to our previous measurements showed a perfect agreement,\cite{RefWorks:6} confirming that $T > \Tn$ as the AFM ordering of the Tm moments significantly affects the VL form factor and suppresses the Pauli paramagnetic effect.\cite{RefWorks:38} Vortex lattices were prepared by cooling through $\Tc$ in a constant field (FC) and, in some cases, followed by a damped small-amplitude field oscillations (FCO). Background measurements were measured at 14~K and were subtracted from the foreground measurements. The diffracted neutrons were detected by a two-dimensional $^3$He position-sensitive proportional counter.

\section{Results}
\begin{figure*}
 \includegraphics{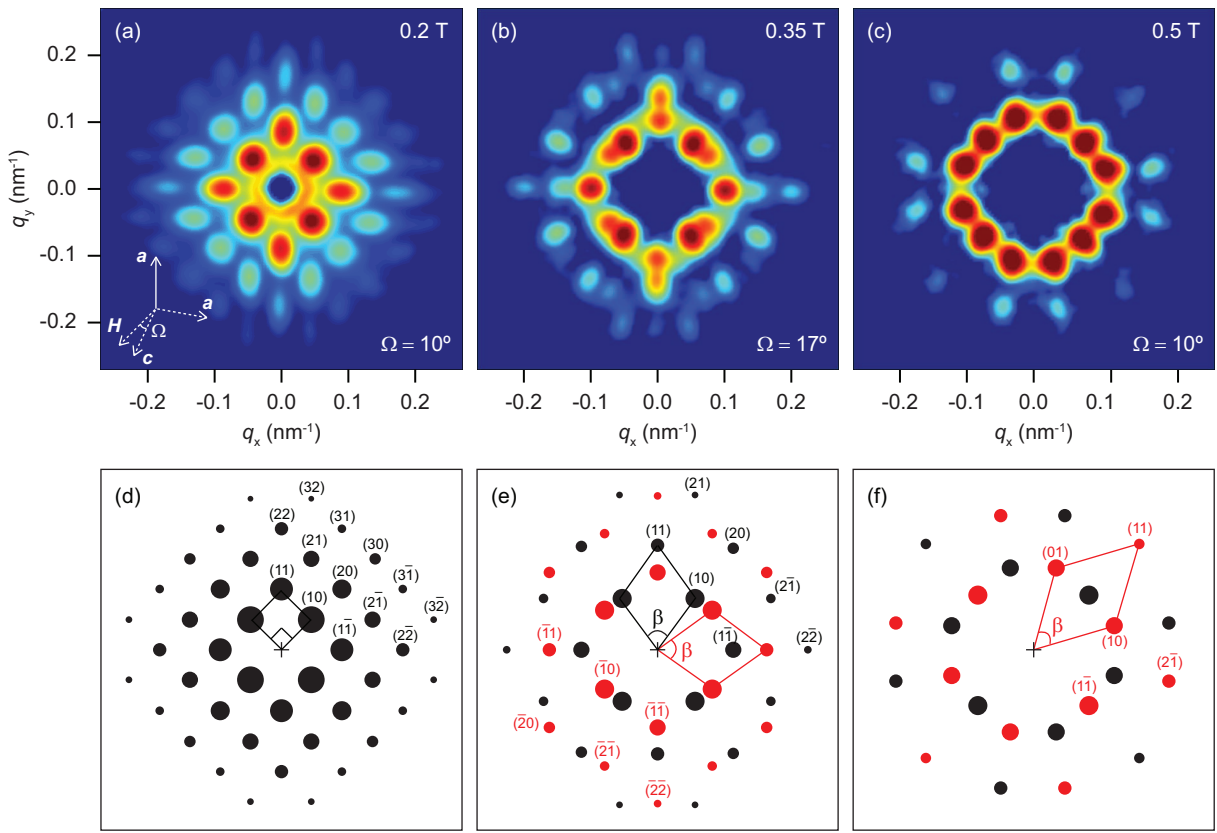}
 \caption{\label{Fig1}
          (Color online)
          SANS diffraction patterns showing the square, rhombic and hexagonal VL phases in \Tm \ at $1.6$~K for applied fields of $0.2$~T (a), $0.35$~T (b) and $0.5$~T (c) respectively. The images are obtained by summing measurements at multiple rotation and tilt angles to satisfy the Bragg condition for the different reflections. The scattered intensity is shown on a logarithmic false color scale to make strong and weak reflections simultaneously visible. The orientation of the crystalline axes and the magnetic field is shown in the inset to panel (a), where $\Omega$ is the angle between the field and the $c$ axis. The indexing of the VL Bragg peaks are shown for one quadrant in the schematics in panels (d) to (f)  with the size of the circles indicating the intensity. For the rhombic and hexagonal VL phases two domain orientations with an opening angle $\beta$ are observed, as shown by the black and red circles in (e) and (f). With increasing field the VL Bragg reflections move to longer scattering vectors $q$ and decrease in intensity, making fewer of them visible.}
\end{figure*}

Examples of the square, and the high-field rhombic and hexagonal VL phases observed in \Tm \ are shown in Fig.~\ref{Fig1}.
In this material the low-field hexagonal and rhombic lattices occur below the convenient field range for SANS experiments. However, a hexagonal VL have been observed at 2~mT by Bitter decoration.\cite{RefWorks:38,AbrahamsenThesis} For all three cases shown in Fig.~\ref{Fig1} the measurements were extended to include as many higher-order reflections as possible within reasonable count times. The square VL diffraction pattern in panel~(a) was obtained at $0.2$~T and shows Bragg reflections with scattering vectors given by $q_{hk} = (h^2 + k^2)^{1/2} \, q_0$, where $q_0 = 2\pi(B/\phi_0)^{1/2}$ and $\phi_0 = h/2e = 2068$~T~nm$^2$ is the flux quantum. Fig.~\ref{Fig1}(d) shows the indexing of the peaks in one quadrant, with the remaining obtainable by symmetry. The diffraction pattern in Fig.~\ref{Fig1}(a) was obtained with $\Omega = 10^\circ$. Due to the very modest $ac$-anisotropy of the extrapolated orbital upper critical field $\Gamma = H_{\text{c2,orb}}^{\perp c}/H_{\text{c2,orb}}^{\parallel c} \approx 1.2$ in \Tm,\cite{RefWorks:807} no distortion of the VL due to the rotation of the field away from the $c$ axis is detectable.

Increasing the field to $0.35$~T causes the VL to change to a two-domain rhombic structure as shown in Fig.~\ref{Fig1}(b) with the corresponding indexing in panel~(e). The rhombic VL has an opening angle $\beta = 73.0^\circ$ and scattering vectors $q_{hk} = (h^2 + k^2 + 2hk \cos \beta)^{1/2} \, q_0$, where $q_0 = 2\pi(B/\phi_0 \sin \beta)^{1/2}$. These measurements were done with $\Omega = 17^\circ$, chosen to favor one of the two VL domain orientations while at the same time keeping the distortion due to the $ac$-anisotropy small. For this $\Omega$ the minority domain (red circles) is sufficiently suppressed to allow a reliable measurement of the intensity of the higher order majority domain (black circles) reflections. No measurable difference in the VL opening angle $\beta$ was observed between the two domains.

Finally, as the field is increased to $0.5$~T, a distorted hexagonal VL was observed as shown in Fig.~\ref{Fig1}(c) with the indexing in panel~(f), and with an opening angle $\beta =  56.1^\circ$. The magnitude of the scattering vector is given by the same expression as in the rhombic phase. Note that equivalent scattering vectors were chosen as the unit vectors, leading to $q_{1 \bar{1}}$ being slightly shorter than $q_{10}$ as $\beta < 60^\circ$. However, this is merely a naming convention and will not affect the analysis of the scattered intensity. The measurements were performed at a field rotation $\Omega = 10^\circ$. For the hexagonal VL orientation the two domains are orientated equivalently with respect to the field rotation axis and are thus equally populated. A distorted hexagonal VL was also observed at $0.6$~T (not shown) with an opening angle $\beta =  56.3^\circ$. However, at this field, no higher-order peaks were measurable due to the decreasing scattered intensity with increasing scattering vector and applied field.

\begin{figure}
 \includegraphics{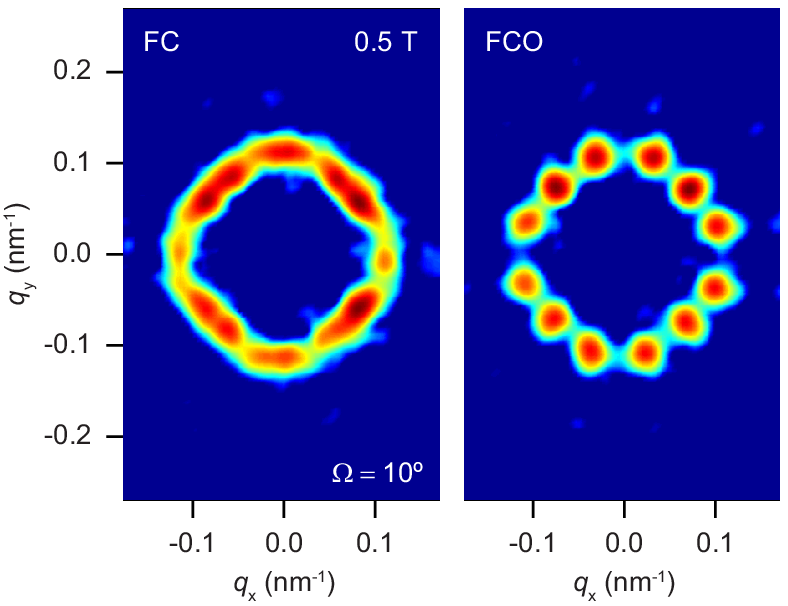}
 \caption{\label{Fig2}
          (Color online)
          Comparison of VLs at $0.5$~T ($\Omega = 10^\circ$) and $1.6$~K. The diffraction patterns were obtained directly following a field-cooling procedure (FC) and after the application of a damped field oscillation with an initial amplitude of 25~mT (FCO). No higher order VL reflections were observed due to shorter count times as compared to Fig.~\ref{Fig1}(c).}
\end{figure}

The diffraction patterns in Fig.~\ref{Fig1} were obtained following a preparation method chosen to produce the most ordered VL.
Depending on the level and strength of vortex pinning in the host material relative to the vortex-vortex interaction, the optimal preparation may either be a field cooling (FC) procedure or a FC followed by a damped small-amplitude field oscillation (FCO).\cite{RefWorks:795} While we found no difference between VLs prepared by the two different methods at $0.2$~T, the FCO procedure provided a substantially better ordered VL at higher fields. This is seen in Fig.~\ref{Fig2}, which compares FC and FCO VL diffraction patterns obtained at $0.5$~T.
It is likely that the VL disordering observed above $0.2$~T in the FC case is due to the crossing of VL symmetry phase transitions while cooling from $T > \Tc$ to the measurement temperature of $1.6$~K.\cite{RefWorks:38,AbrahamsenThesis} Based on these findings, measurements were performed following a FC at $0.2$~T and following a FCO for $0.35$~T and above. In all cases where a FCO procedure was used, the initial amplitude of the damped field oscillation was 5\% of the final field.

\begin{figure}
 \includegraphics{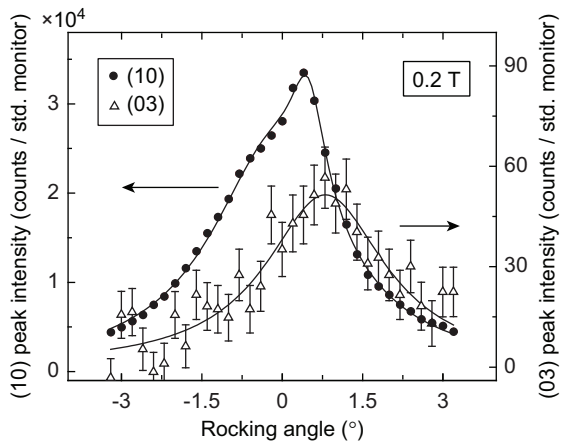}
 \caption{\label{Fig3}
          Rocking curves for the square VL (10) and (03) Bragg reflections at $0.2$~T and $1.6$~K corresponding to Fig.~\ref{Fig1}(a) and (d). Note the different axes for the two reflections. Each angle was counted for 9~min. For the (10) reflection the error bars are smaller than the symbol size. The (10) reflection is fitted with a double Lorentzian function due to the irregular shape with a shoulder left of the main peak. The (03) reflections is fitted by a single Lorentzian.}
\end{figure}

We now turn to measurements of the VL form factors which are the main focus of this study. The form factor $F(q_{hk})$ is the Fourier transform, at scattering wave vector $q_{hk}$, of the two-dimensional magnetic field modulation due to the VL. It is related to the integrated reflectivity $R$, which is obtained by rotating and/or tilting the cryomagnet and sample such that the VL scattering vectors cut through the Ewald sphere. Examples of rocking curves obtained in this fashion are shown in Fig.~\ref{Fig3}.
In contrast to other members of the borocarbide superconductors such as \Lu,\cite{RefWorks:3} the rocking curves in \Tm \ are found to be broad and with asymmetric line-shapes, necessitating multi-function fits to accurately obtain the integrated scattered intensity as shown for the (10)-peak. For the higher-order reflections the line-shapes appear more regular, as seen for the (03)-peak, and can be fitted by a single Lorentzian, although this may also be a result of poorer signal-to-noise. Broad VL rocking curves, but with regular line shapes, were also found in other work, using \Tm \ single crystals from a different source.\cite{LevettThesis} It is possible that the present sample have more mosaicity which can explain the asymmetric lineshape. Nonetheless, the current rocking curves are still narrow enough to be easily measurable as shown in Fig.~\ref{Fig3} and thus, the total scattered intensity can be precisely determined for each VL reflection.

With the strong scattering from the VL in \Tm \ it is necessary to consider whether multiple scattering is affecting the measured intensities. This was discussed in detail by Densmore {\em et al.} in the case of \Lu,\cite{RefWorks:3} where it was shown that multiple scattering did not pose a problem. In the case of \Tm, the integrated intensity is even stronger, but since the rocking curve is also significantly broader, the fraction of the incident neutrons scattered by the VL is $\leq 0.4$\%, which is almost identical to \Lu. We thus conclude that the error in the measured intensities due to multiple scattering is insignificant.

\begin{figure}
  \includegraphics{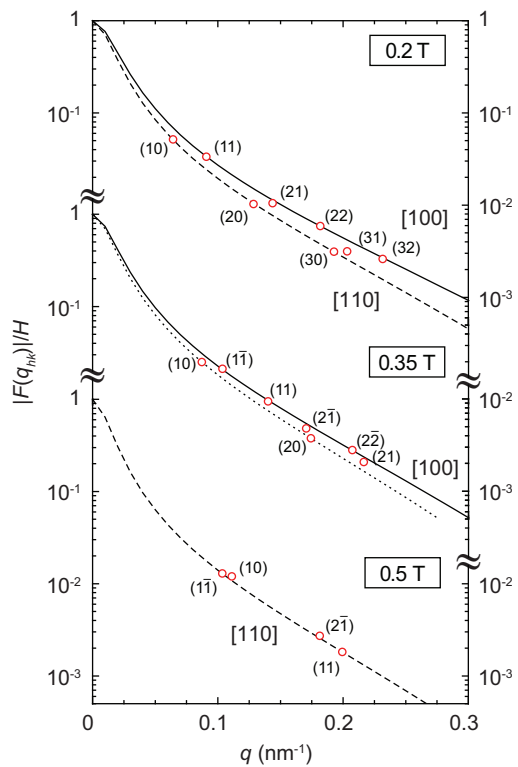}
  \caption{\label{Fig4}
           (Color online)
           VL form factor divided by the applied field versus scattering vector $q$ for all measured reflections at $0.2$, $0.35$ and $0.5$~T. For all fields and reflections the error bars are smaller than the symbol size.
           The curves are fits to the London model as described in the text. Full and dashed lines correspond to VL Bragg peaks along the crystalline [100] and [110] directions respectively.}
\end{figure}

The integrated intensity is divided by the incident neutron flux to yield the integrated reflectivity
\begin{equation}
  R_{hk} = \frac{2\pi \gamma^2 \lambda_{\text n}^2 t}{16 \phi_0^2 q_{hk}} \left| F(q_{hk}) \right| ^2,
\end{equation}
where $\gamma = 1.913$ is the magnetic moment of the neutron in nuclear magnetons and $t$ is the average sample thickness (differences in $t$ due to the change in $\Omega$ are $\leq 3\%$ and thus insignificant). The intensity for each reflection is corrected for the angle at which it cuts the Ewald sphere during the measurement of the rocking curve (Lorentz factor). Fig.~\ref{Fig4} summarizes the measured VL form factors for all reflections and fields of $0.2$, $0.35$ and $0.5$~T.
For the two higher fields the intensities of equivalent reflections for the two domains have been added to obtain scattering from all the vortices in the sample in order to compare it directly to the square VL at $0.2$~T.

\section{Discussion}
We will now discuss how the measured VL form factors can be used to study the evolution of the superconducting basal plane anisotropy in \Tm. Qualitatively, a reduction of the anisotropy with increasing field is directly evident from Fig.~\ref{Fig4}. For the square VL at $0.2$~T the form factors do not fall on a single curve, as expected for an isotropic superconductor, but rather lie on or between two curves going through $|F(q_{h0})|/H$ and $|F(q_{hh})|/H$, respectively. As these two limiting curves are $45^\circ$ apart, their separation is a measure of the four-fold basal plane anisotropy. With increasing field we see that the separations between the form factors along different directions decrease, indicating that the superconducting state becomes more isotropic. However, the situation is complicated by the change in the VL symmetry, which changes the position of the reflections with respect to the crystalline axes.

The curves in Fig.~\ref{Fig4} are fits to the London model, extended by a Gaussian cutoff to take into account the finite vortex core size,\cite{RefWorks:300,RefWorks:3}
\begin{equation}
  F_{\text{L}}(q)=\frac{H}{1+(q \lambda)^2} \; e^{-c(q \xi)^2}
\end{equation}
where $\lambda$ and $\xi$ are, respectively, the penetration depth and the coherence length while $c$ is a constant typically taken to be between $1/4$ and 2.\cite{RefWorks:300}
Using $c = 1/2$, the fitted values of $\lambda$ and $\xi$ for the curves in Fig.~\ref{Fig4}, which correspond to crystalline high-symmetry directions, are given in Table~\ref{lambdaxitable}.
\begin{table}
  \begin{tabular}{c|cc|cc|cc}
             & \multicolumn{2}{c|}{[110]}  & \multicolumn{2}{c|}{[100]}  \\
             & $\lambda$ (nm) & $\xi$ (nm) & $\lambda$ (nm) & $\xi$ (nm) \\
    \hline
    $0.2$~T  & 64.1 & 6.28 & 55.3 & 5.49 \\
    $0.35$~T &      &      & 59.3 & 6.34 \\
    $0.5$~T  & 74.5 & 6.73 &      &      \\
    \hline
  \end{tabular}
  \caption{\label{lambdaxitable}
           Coefficients of London model fits ($c = 0.5$) shown in Fig.~\ref{Fig4} in the case where the VL Bragg peaks are along the crystalline [100] or [110] directions.}
\end{table}
It is important to stress that the primary objective of the fitting is to obtain an analytical expression for the VL form factor for each field and direction, and the coefficients {\em are not to be taken as accurate determination of $\lambda$ and $\xi$}. Nonetheless, Table~\ref{lambdaxitable} shows that increasing the field from $0.2$ to $0.5$~T both the fitted values for the penetration depth and the coherence length along the crystalline [110] direction increases, consistent with a reduction of the field modulation and an expansion of the vortex core. The same is seen for $\lambda$ and $\xi$ along $[100]$ as the field is increased from $0.2$ to $0.35$~T.

At $0.2$~T, a measure of the superconducting basal plane anisotropy is obtained from the ratio $(\lambda_{110}/\lambda_{100})^2 = 1.34$. We note that a very similar value is found for the coherence length ratio $(\xi_{110}/\xi_{100})^2 = 1.31$. At the higher fields however, the changing VL symmetry causes the reflections to move. Notably, at the two higher fields there are only VL Bragg peaks along one of the two crystalline high-symmetry directions ([100] for the rhombic VL at $0.35$~T; [110] for the hexagonal VL at $0.5$~T). As a result it is not possible to directly extract the superconducting basal plane anisotropy at the higher fields. In the following we present a more careful analysis of the field dependence of the anisotropy.

\begin{figure}
  \includegraphics{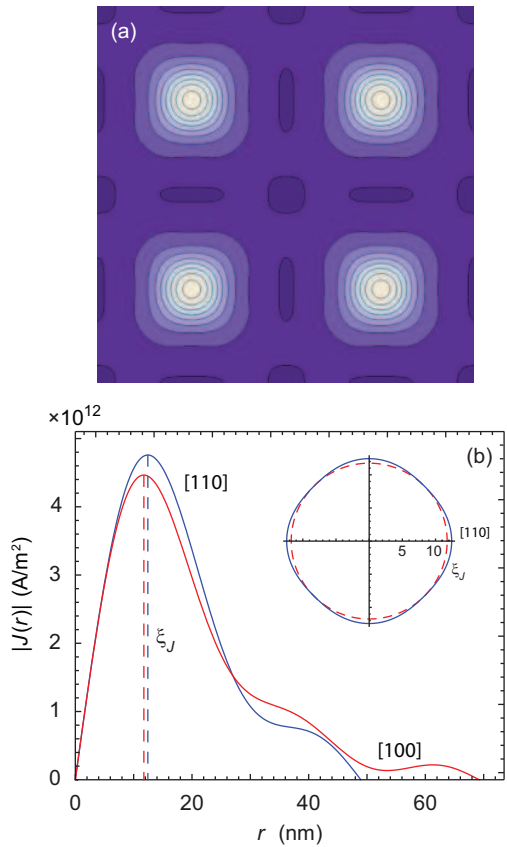}
  \caption{\label{Fig5}
           (Color online)
           Real space magnetic field reconstruction from the measured VL form factors at $\mu_0 H = 0.2$~T and $T = 1.6$~K.
           (a) Contour plot of the magnetic field, showing four VL unit cells with a vortex spacing of 98~nm, corresponding to a magnetic induction $B = 0.216$~T obtained from the magnitude of the scattering vector. Note that the image is rotated $45^\circ$ with respect to Fig.~\ref{Fig1} such that $\{110\}$ directions are horizontal/vertical. The lowest contour corresponds to $B = 198$~mT and the contour spacing is 15~mT.
           (b) Current density as a function of distance from the vortex center along the VL nearest-neighbor direction ([110]) and the unit cell diagonal ([100]). The inset shows the value of $\xi_J$ (distance of maximum current) in the basal plane. To emphasize the fourfold anisotropy, a circle with radius $\xi^{100}_J$ is shown by the dashed line.}
\end{figure}

As shown by Densmore {\em et al.},\cite{RefWorks:3} a conceptually simple and model-independent method to obtain a measure of the basal plane anisotropy is by a real space magnetic field reconstruction using
\begin{equation}
  B(\bm{r}) = \sum_{hk} F(\bm{q}_{hk}) \: e^{i \bm{q}_{hk} \cdot \bm{r}}.
  \label{Bsum}
\end{equation}
Since the SANS measurements only measure the absolute magnitude of the form factors, this requires an assumption about the relative sign of the Fourier components $F(\bm{q}_{hk})$. In the case of \Lu, a comparison to muon spin-rotation measurements showed that for fields below $\sim \Hcii/3$ the form factors all have the same sign,\cite{RefWorks:3} in agreement with the prediction of the London model as well as numerical results based on the Eilenberger equations.\cite{RefWorks:267} In the case of \Tm \ the measured $\Hcii = 0.75$~T at $1.6$~K is severely Pauli limited.\cite{RefWorks:869} Instead, we use the extrapolated orbital upper critical field $H_{\text{c2,orb}} = 4.3$~T,\cite{RefWorks:807} yielding an estimated upper limit of $1.4$~T for the all-equal sign scheme. This is thus expected to be valid for all the measurements in this report. Fig.~\ref{Fig5}(a) shows the real space field reconstruction obtained from the measured form factors in an applied field of $0.2$~T.
The accuracy of the reconstruction depends on the number of reflections included in the sum in Eq.~(\ref{Bsum}). In the present case, the magnitudes of the form factors for the (30), (31) and (32)-reflections, which were the highest order peaks which could be measured, are less than 1~mT and do not change the reconstruction to any significant degree. We note that the vortex spacing $d = 2\pi/q_{10}$ can be determined from the magnitude of the VL scattering vector, which allows for a determination of the magnetic induction $B = \phi_0/d^2 = 0.216$~T in good agreement with earlier reports.\cite{RefWorks:6} That $B > \mu_0 H$ is due to the paramagnetism of \Tm \ for $T > \Tn$. The magnitude of the field modulation $\approx 130$~mT or $0.6 B$ is much larger than the 10\% observed in non-magnetic \Lu \ in an applied field of $0.5$~T. Extrapolating the \Lu \ form factors to a field of $0.266$~T yields a modest estimate for the increase of the field modulation:
$\exp[2\pi^2 (8.22 \mbox{~nm})^2 \, (0.5 \mbox{~T} - 0.266 \mbox{~T})/\phi_0] = 1.16$.\cite{RefWorks:3} This is still much smaller than the 60\% observed here for \Tm. This difference is a manifestation of the strong Pauli paramagnetic effects, which leads to a significant polarization of the unpaired quasi-particle spins in the vortex cores of \Tm \ and creates a periodic magnetization that adds significantly to the field modulation from the circulating supercurrents.

From the field reconstruction, one can calculate the current distribution by $\mu_0 \bm{J} = \nabla \times \bm{B}$, which contains contributions from {\em both} the supercurrents and the periodic magnetization and which can not be easily deconvoluted.\cite{RefWorks:756,RefWorks:804} Fig.~\ref{Fig5}(b) shows $|\bm{J}(\bm{r})|$ as a function of the distance from the vortex core along the VL nearest neighbor direction ([100]) and the unit cell diagonal ([110]). The distance from the vortex center to the peak of the current density provides a measure of the core size $\xi_J$,\cite{RefWorks:267,RefWorks:63} which is seen to differ for the two directions shown. The inset to Fig.~\ref{Fig5}(b) shows the vortex core size in the plane perpendicular to the field, which displays a clear four-fold anisotropy. Specifically, we find $\xi^{110}_J = 12.4$~nm and $\xi^{100}_J = 11.7$~nm. The ratio between these two values is $1.06$, slightly smaller than the $1.08$ found for the square VL in \Lu.\cite{RefWorks:3} Since the VL at $0.2$~T is close to the onset of the square-to-rhombic transition in \Tm, the measured anisotropy provides an estimate of the critical value necessary for stabilizing a square symmetry. Comparing the values for $\xi_J$ to the London model fits at $0.2$~T listed in Table~\ref{lambdaxitable}, one finds a substantial difference, unlike our earlier measurements on \Lu \ where the two were found to be in excellent agreement.\cite{RefWorks:3} A theoretical analysis of the interdependence of the superconducting and magnetic properties of \Tm \ by Jensen and Hedeg\aa rd gives an estimate of the orbital critical field $H_{\text{c2,orb}} = 4.3$~T at $1.6$~K,\cite{RefWorks:807} yielding a zero-field coherence length of $8.75$~nm. However, this ignores the contribution to the core size from the spin-polarized quasi-particles in the vortex core and we thus expect that $\xi_J$ obtained from the field reconstruction to be a more accurate measure of the actual vortex core size. It should also be noted that the anisotropy of $\xi_J$ is smaller than the ratio from Table~\ref{lambdaxitable}, again illustrating that while $\xi$ and $\xi_J$ are related they are not identical.

\begin{figure}
  \includegraphics{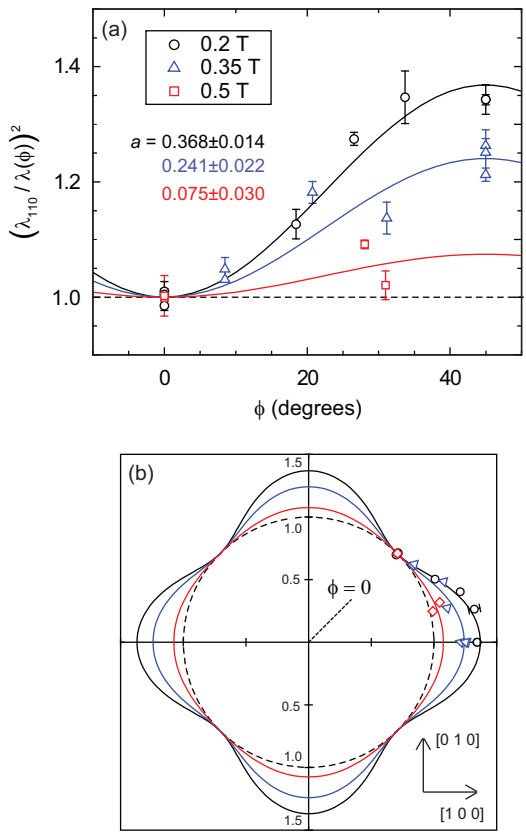}
  \caption{\label{Fig6}
           (Color online)
           (a) Angle dependence of the superconducting basal plane anisotropy ratio calculated from the VL form factors as described in the text. For each field the curves show a fit to the anisotropy function given in Eq.~(\ref{anisotropyfct}).
           (b) The same results in polar coordinates.}
\end{figure}

While a field reconstruction at higher fields would be desirable, it is not possible to measure enough higher-order reflections necessary to obtain this with satisfactory accuracy. This is the case particularly in the hexagonal VL phase. Instead we return to the London model fits in Fig.~\ref{Fig4} to obtain a quantitative, symmetry independent measure of the field dependence of the superconducting basal plane anisotropy in \Tm. The fits indicate that the VL form factor along any direction in the basal plane may be parameterized by
\begin{equation}
  \frac{F(\phi)}{H} =\frac{1}{1+[q \, \lambda(\phi)]^2} \; e^{-c[q \, \xi(\phi)]^2},
  \label{FFvsphi}
\end{equation}
where $\phi$ is the angle relative to the crystalline [110]-direction. This yields an expression for the $\phi$-dependence of $\lambda$ and thus an angle dependent anisotropy ratio
\begin{equation}
  \left( \frac{\lambda_{110}}{\lambda(\phi)} \right)^2
    = \frac{\lambda_{110}^2 \, q^2 \, F(\phi)}{e^{-c [q \, \xi(\phi)]^2} - F(\phi)}.
  \label{anisotropyratio}
\end{equation}
This ratio is expected to exhibit a four-fold symmetry
\begin{equation}
  \left( \frac{\lambda_{110}}{\lambda(\phi)} \right)^2 = 1 + a \frac{1 - \cos 4\phi}{2},
  \label{anisotropyfct}
\end{equation}
where the parameter $a$ is the anisotropy amplitude. As noted earlier, the fitted values for $\lambda$ and $\xi$ given in Table~\ref{lambdaxitable} show $(\lambda_{110}/\lambda_{100})^2 \approx (\xi_{110}/\xi_{100})^2$, indicating that $\lambda(\phi)$ and $\xi(\phi)$ have the same anisotropy amplitude. From Eq.~(\ref{anisotropyfct}) we obtain
\begin{multline}
  \frac{1}{[\lambda,\xi]^2(\phi)} = \\
    \frac{1}{[\lambda,\xi]_{110}^2} + \left( \frac{1}{[\lambda,\xi]_{100}^2} - \frac{1}{[\lambda,\xi]_{110}^2} \right) \frac{1 - \cos 4\phi}{2}.
  \label{lambdaxiphi}
\end{multline}
In Fig.~\ref{Fig6} we show the anisotropy ratio obtained by Eq.~(\ref{anisotropyratio}) using the measured VL form factors at $0.2$~T and $\xi(\phi)$ calculated using Eq.~(\ref{lambdaxiphi}) and the values in Table~\ref{lambdaxitable}.
The data are well fitted by Eq.~(\ref{anisotropyfct}) yielding an anisotropy amplitude $a = 0.368 \pm 0.014$ in good agreement with the result based solely on the form factors corresponding the VL Bragg peaks on the high symmetry [110] and [100] directions (Table~\ref{lambdaxitable}).

Having demonstrated that the approach above yields a consistent results for the anisotropy we now apply it to the $0.35$ and $0.5$~T measurements. Here we simultaneously adjust the value of $a$ used to calculate $\xi(\phi)$ in Eq.~(\ref{anisotropyratio}) and the fitted value obtained by Eq.~(\ref{anisotropyfct}) to ensure a self consistent result. This approach allows a determination of the anisotropy amplitude even in cases where there are only VL Bragg reflections along a single crystalline high symmetry direction ([110] or [100]). Figure~\ref{Fig6} shows the results of this analysis, providing a quantitative measure of the monotonically decreasing superconducting basal plane anisotropy with increasing field. This is in stark contrast to the behavior found for non-magnetic \Lu, where a similar analysis on the data from from Ref.~\onlinecite{RefWorks:3} yields $a = 0.456 \pm 0.060$ at $0.5$~T increasing slightly to $0.492 \pm 0.008$ at $1.0$~T.

Our results provide a direct confirmation of theoretical predictions that in superconductors with strong Pauli paramagnetic effects, paramagnetic depairing causes the vortex cores to expand and also become more isotropic as one approach $\Hcii$.\cite{IchiokaPC,RefWorks:867} One would therefore also expect that the anisotropy of {\Tm} will increase below the antiferromagnetic ordering at $\Tn = 1.5$~K where the Pauli paramagnetic effects are known to decrease.\cite{RefWorks:38}

\section{Conclusion}
In summary, we have investigated the field dependence of the superconducting four-fold basal plane anisotropy of \Tm. We have observed and quantified the decreasing anisotropy with increasing applied field, which provides an explanation of the reentrant square VL phase. The decreasing anisotropy is attributed due to the strong Pauli paramagnetic effects observed in \Tm, leading to an expansion of the vortex cores near $\Hcii$. We believe that a similar mechanism is responsible for the reentrance of the square VL phase observed in \Ce.

\section*{Acknowledgements}
We acknowledge discussions with E. M. Forgan, M. Ichioka, K. Machida, V. P. Michal, V. P. Mineev, and J. S. White.
This work was supported by the US National Science Foundation through grant DMR-0804887.
J.M.D. was supported in part by an appointment to the U.S. Army Research Laboratory Postdoctoral Fellowship Program administered by the Oak Ridge Associated Universities through a contract with the U.S. Army Research Laboratory.
K.J.S. recognizes support from the Notre Dame Institute for Scholarship in the Liberal Arts.
M.L. acknowledges support from DanScatt.
We acknowledge the support of the National Institute of Standards and Technology, U.S. Department of Commerce, in providing the neutron research facilities used in this work.
Research at Oak Ridge National Laboratory's High Flux Isotope Reactor was sponsored by the Scientific User Facilities Division, Office of Basic Energy Sciences, U. S. Department of Energy.
Work at Ames Laboratory was supported by the U.S. Department of Energy, Office of Basic Energy Science, Division of Materials Sciences and Engineering. Ames Laboratory is operated for the U.S. Department of Energy by Iowa State University under Contract No. DE-AC02-07CH11358.

\bibliographystyle{apsrev4-1}

\begin{thebibliography}{99}
\bibitem{RefWorks:751}
  V. G. Kogan, M. Bullock, B. Harmon, P. Miranovi\'{c}, L. Dobrosavljevi\'{c}-Gruji\'{c}, P. L. Gammel, and D. J. Bishop,
  Phys. Rev. B \textbf{55}, R8693 (1997).
\bibitem{RefWorks:754}
  K. M. Suzuki, K. Inoue, P. Miranovic, M. Ichioka, and K. Machida,
  J. Phys. Soc. Japan \textbf{79}, 013702 (2010).
\bibitem{RefWorks:38}
  M. R. Eskildsen, K. Harada, P. L. Gammel, A. B. Abrahamsen, N. H. Andersen, G. Ernst, A. P. Ramirez, D. J. Bishop, K. Mortensen, D. G. Naugle, K. D. D. Rathnayaka, and P. C. Canfield,
  Nature \textbf{393}, 242 (1998).
\bibitem{RefWorks:22}
  M. R. Eskildsen, C. D. Dewhurst, B. W. Hoogenboom, C. Petrovic, P. C. Canfield,
  Phys. Rev. Lett. \textbf{90}, 187001 (2003).
\bibitem{RefWorks:5}
  A. D. Bianchi, M. Kenzelmann, L. DeBeer-Schmitt, J. S. White, E. M. Forgan, J. Mesot, M. Zolliker, J. Kohlbrecher, R. Movshovich, E. D. Bauer, J. L. Sarrao, Z. Fisk, C. Petrovic, and M. R. Eskildsen,
  Science \textbf{319}, 177 (2008).
\bibitem{RefWorks:331}
  S. Ohira-Kawamura, H. Shishido, H. Kawano-Furukawa, B. Lake, A. Wiedenmann, K. Kiefer, T. Shibauchi, and Y. Matsuda,
  J. Phys. Soc. Japan \textbf{77}, 023702 (2008).
\bibitem{RefWorks:1}
  J. S. White, P. Das, M. R. Eskildsen, L. DeBeer-Schmitt, E. M. Forgan, A. D. Bianchi, M. Kenzelmann, M. Zolliker, S. Gerber, J. L. Gavilano, J. Mesot, R. Movshovich, E. D. Bauer, J. L. Sarrao, and C. Petrovic,
  New J. Phys. \textbf{12}, 023026 (2010).
\bibitem{RefWorks:8}
  L. DeBeer-Schmitt, C. D. Dewhurst, B. W. Hoogenboom, C. Petrovic, and M. R. Eskildsen,
  Phys. Rev. Lett. \textbf{97}, 127001 (2006).
\bibitem{RefWorks:6}
  L. DeBeer-Schmitt, M. R. Eskildsen, M. Ichioka, K. Machida, N. Jenkins, C. D. Dewhurst, A. B. Abrahamsen, S. L. Bud'ko, and P. C. Canfield,
  Phys. Rev. Lett. \textbf{99}, 167001 (2007).
\bibitem{RefWorks:756}
  M. Ichioka and K. Machida,
  Phys. Rev. B \textbf{76}, 064502 (2007).
\bibitem{RefWorks:805}
  R. Ikeda, Y. Hatakeyama and K. Aoyama,
  Phys. Rev. B \textbf{82}, 060510 (2010).
\bibitem{RefWorks:804}
  V. P. Michal and V. P. Mineev,
  Phys. Rev. B \textbf{82}, 104505 (2010).
\bibitem{Aoyama}
  K. Aoyama and R. Ikeda,
  Phys. Rev. B \textbf{84}, 184516 (2011).
\bibitem{IchiokaPC}
  M. Ichioka and K. Machida,
  Private communication.
\bibitem{RefWorks:867}
  N. Hiasa and R. Ikeda,
  Phys. Rev. Lett. \textbf{101}, 027001 (2008).
\bibitem{RefWorks:3}
  J. M. Densmore, P. Das, K. Rovira, T. D. Blasius, L. DeBeer-Schmitt, N. Jenkins, D. McK Paul, C. D. Dewhurst, S. L. Bud'ko, P. C. Canfield, and M. R. Eskildsen,
  Phys. Rev. B \textbf{79}, 174522 (2009).
\bibitem{RefWorks:397}
  P. C. Canfield and I. R. Fisher,
  J. Cryst. Growth \textbf{225}, 155 (2001).
\bibitem{RefWorks:269}
  B. K. Cho, M. Xu, P. C. Canfield, L. L. Miller, and D. C. Johnston,
  Phys. Rev. B \textbf{52}, 3676 (1995).
\bibitem{RefWorks:870}
  L. J. Chang, C. V. Tomy, D. M. Paul, and C. Ritter,
  Phys. Rev. B \textbf{54}, 9031 (1996).
\bibitem{RefWorks:463}
  B. Sternlieb, C. Stassis, A. I. Goldman, P. C. Canfield, and S. Shapiro,
  J. Appl. Phys. \textbf{81}, 4937 (1997).
\bibitem{RefWorks:868}
  J. W. Lynn, S. Skanthakumar, Q. Huang, S. K. Sinha, Z. Hossain, L. C. Gupta, R. Nagarajan, and C. Godart,
  Phys. Rev. B \textbf{55}, 6584 (1997).
\bibitem{AbrahamsenThesis}
  A. B. Abrahamsen,
  {\em Possible magnetism in vortex cores of superconducting TmNi$_2$B$_2$C studied by small angle neutron scattering},
  Ph. D. thesis, Danish Technical University (2003), http://www.risoe.dtu.dk/rispubl/AFM/afmpdf/ris-r-1425.pdf.
\bibitem{RefWorks:807}
  J. Jensen and P. Hedeg{\aa}rd,
  Phys. Rev. B \textbf{76}, 094504 (2007).
\bibitem{RefWorks:795}
  C. D. Dewhurst, S. J. Levett, and D. M. Paul,
  Phys. Rev. B \textbf{72}, 014542 (2005).
\bibitem{LevettThesis}
  S. J. Levett,
  {\em Flux-line Ordering in Low-$T_c$ Superconductors},
  Ph. D. thesis, University of Warwick (2003).
\bibitem{RefWorks:300}
  A. Yaouanc, P. Dalmas de R\'{e}otier, and E. H. Brandt,
  Phys. Rev. B \textbf{55}, 11107 (1997).
\bibitem{RefWorks:267}
  M. Ichioka, A. Hasegawa, and K. Machida,
  Phys. Rev. B \textbf{59}, 8902 (1999).
\bibitem{RefWorks:869}
  D. G. Naugle, K. D. D. Rathnayaka, K. Clark, and P. C. Canfield,
  Int. J. Mod. Phys. B \textbf{13}, 3715 (1999).
\bibitem{RefWorks:63}
  J. E. Sonier,
  J. Phys. Condens. Matter \textbf{16}, S4499 (2004).
\end{thebibliography}

\newpage

\end{document}